\documentclass[pdflatex,sn-mathphys-ay]{article}

\usepackage{natbib}
\usepackage{graphicx}%
\usepackage{multirow}%
\usepackage{amsmath,amssymb,amsfonts}%
\usepackage{amsthm}%
\usepackage{mathrsfs}%
\usepackage[title]{appendix}%
\usepackage{xcolor}%
\usepackage{textcomp}%
\usepackage{manyfoot}%
\usepackage{booktabs}%
\usepackage{algorithm}%
\usepackage{algorithmicx}%
\usepackage{algpseudocode}%
\usepackage{listings}
\usepackage{bm}
\usepackage{lscape} 
\usepackage{hyperref}
\title{Clustering Methods for Identifying and Modelling Areas with Similar Temperature Variations}
\author{Edoardo Otranto\\
	Department of Social Sciences and Economics\\
	Sapienza University of Rome\\
	Piazzale Aldo Moro, 5; 00185 Rome\\
	e-mail: edoardo.otranto@uniroma1.it}

\begin{document}


\maketitle


\begin{abstract}
This paper proposes a novel data-driven approach for identifying and modelling areas with similar temperature variations throufigureh clustering and Space-Time AutoRegressive (STAR) models. Using annual temperature data from 168 countries (1901-2022), we apply three clustering methods based on (i) warming rates, (ii) annual temperature variations, and (iii) persistence of variation signs, using Euclidean and Hamming distances. These clusters are then employed to construct alternative spatial weight matrices for STAR models. Empirical results show that distance-based STAR models outperform classical contiguity-based ones, both in-sample and out-of-sample, with the Hamming distance-based STAR model achieving the best predictive accuracy. The study demonstrates that using statistical similarity rather than geographical proximity improves the modelling of global temperature dynamics, suggesting broader applicability to other environmental and socioeconomic datasets.
\end{abstract}

\thanks{
	{\bf Acknowledgements}
	The author acknowledges financial support from the Italian PRIN 2022 grant "Methodological and computational issues in large-scale time series models for economics and finance" (20223725WE), funded by the European Union - Next Generation EU, Mission 4 Component  1 CUP 2J53D23003960006.
	}
	
\section{Introduction}
\label{intro}
Temperature change varies across countries for several reasons, including greenhouse gas emissions and geographic location. Consequently, European countries exhibit a more pronounced warming trend than those in other continents, while North America-and northern regions in general-have experienced strong temperature increases in recent years due to the Arctic amplification effect \citep{Previdi_Smith_Polvani:2021}.

Several studies have also shown that Asia and Africa have undergone substantial temperature increases, although not as pronounced as in Europe or North America, whereas Oceania exhibits the lowest warming. For example, the Food and Agriculture Organization of the United Nations report \citep{FAO_2025}, which covers 198 countries and 39 territories, provides observed changes in land surface temperatures relative to the 1951-1980 baseline, showing that several countries have experienced increases of up to 2$^\circ$C. Similarly, \citet{Shen_etal:2022} analyzed temperature trends in 146 major countries from 1980 to 2019, providing clear evidence of global surface warming with marked spatial heterogeneity and polar amplification. They found that more than 80\% of global land areas have undergone significant warming, occurring more rapidly in the Northern Hemisphere than in the Southern Hemisphere. The most intense warming has been recorded in high northern latitudes, confirming Arctic amplification, while warming rates decrease toward the equator, becoming minimal or even negative. Regions with the strongest warming include Greenland (0.654$^\circ$C per decade), Russia, Ukraine, Eastern Europe, Northern Canada, Northern Africa, and the Middle East, whereas areas with the weakest warming (less than 0.15$^\circ$C per decade) include New Zealand, equatorial South America, Southeast Asia, and Southern Africa.

\citet{Matthews_Graham_Keverian:2014} examined national contributions to global temperature change through emissions, land use, and other drivers, while \citet{Ward_Mahowald:2014} demonstrated how future warming is linked to both past and ongoing emissions. At more localized scales, \citet{Caol_etal:2017} analyzed long-term observational records (1901-2015) for China, identifying persistent warming patterns; \citet{Ruosteenoja_Raisanen:2021} applied 28 global climate models to study monthly and annual mean temperatures in Finland from 1901 to 2018, successfully reproducing the observed warming; and \citet{Gil-Alana_Sauci:2019} used station data from 12 European countries to detect statistically significant warming trends, national variations, and microclimate effects.

Studying temperature trends, identifying clusters of areas with similar thermal dynamics, and developing models that account for these spatial and temporal characteristics are crucial for assessing the impacts of climate change in multiple domains, including macroeconomic \citep{Bilal_Kanzig:2025}, financial \citep{Naifar:2024}, and migration dynamics \citep{McLeman_Smit:2006}.

The natural framework for analyzing temperature change is space-time modelling.\footnote{For a review of these models in environmental contexts, see \citet{Le_Zidek:2006}.} Since the seminal work by \citet{Cliff_Ord:1975}, space-time models have been successfully extended to the ARIMA class \citep{Pfeifer_Deutsch:1980}. Among these, the simple Space-Time AutoRegressive (STAR) model is of particular importance due to its linear structure, straightforward estimation procedure, and several desirable properties, such as its long-run steady-state equilibrium implications \citep{LeSage_Pace:2009}.

A critical component of the STAR model is the specification of the spatial weight matrix, which determines how spatial interactions among units are represented. Typically, such weights are constructed using either contiguity or distance-based criteria \citep{Anselin:2013}. The former approach is particularly well suited to climate studies, given the tendency of neighboring territories to exhibit similar thermal behavior.

In this paper, we analyze a large annual dataset (1901-2022) covering 168 countries, with two main objectives. First, we identify groups of countries that are similar in terms of warming rate, annual temperature variation, and the sign of change using classical clustering methods applied to various distance matrices. Second, we construct spatial weight matrices derived from these distances and clusters to estimate alternative STAR models, which we then compare to a benchmark model based on contiguity-based weights. We demonstrate that the new distance-based specifications achieve superior in-sample and out-of-sample performance.

The paper is organized as follows. Section \ref{sec:data} briefly describes the dataset; Section \ref{sec:cluster} presents three alternative clustering approaches for the countries considered in this study, based on similarity in annual warming rates (Subsection \ref{sec:slope}), similarity in temperature variations (Subsection \ref{sec:diff}), and similarity in the sign of temperature variations (Subsection \ref{sec:sign}). These clusters and their associated distance measures are then used to extend the STAR model specification, whose performance is compared with that of the baseline contiguity-based STAR model in both in-sample and out-of-sample contexts (Section \ref{sec:star}). Section \ref{sec:fin} concludes with some final remarks. The final Appendix provides the complete list of the countries analyzed, with the corresponding cluster to which they belong according to the three alternative criteria illustrated in Section \ref{sec:cluster}.

\begin{table}[t]
	\centering
	\begin{tabular}{l r r r}
		\multicolumn{1}{c}{\bf{Zone}}&\multicolumn{1}{c}{\bf{\# Countries}}&\multicolumn{1}{c}{\bf{\% Sample}}&\multicolumn{1}{c}{\bf{\% Zone}}\\
		Europe&	37&	4.79&93.78\\
		Asia&	38&		17.18&60.21\\
		Eurasia&	5&	15.40&100\\
		Africa&	46&		21.99&94.94\\
		North America&3&	17.66&90.52
		\\
		Central America&	14&	0.61&95.04\\
		South America&	13&		15.12&99.53\\
		Oceania&	12&	7.25&99.70\\
		Global&168&100&87.10\\
	\end{tabular}
	\caption{Distribution of 168 countries across geographical zones: number of countries (Column \emph{\# Countries}), percentage of area covered by the countries in the sample (\emph{\% Sample}), and percentage of global area covered by the countries in the sample compared to the corresponding zone sample (\emph{\% Zone}).}\label{tab:descr}
\end{table}

\section{The Dataset}\label{sec:data}
The dataset used in this study is freely available on Kaggle.\footnote{The dataset can be downloaded from {https://www.kaggle.com/datasets/palinatx/mean-temperature-for-countries-by-year-2014-2022}. It was scraped from the World Bank Climate Knowledge Portal ({https://climateknowledgeportal.worldbank.org/}) and includes data for all available countries from 1901 to 2022. Unfortunately, data for some major countries, such as China and India, are not available for this period.} It contains annual average temperature series for 168 countries spanning the years 1901-2022, for a total of 20,496 observations.

Table~\ref{tab:descr} summarizes the distribution of countries across the main geographical regions. The dataset includes a varying number of countries per region, although the emphasis is naturally placed on the total area covered by the included states. The third column of the table reports the percentage of the sample area represented by each region. With the exception of Asia, all regions cover nearly the entire geographic area (last column), with percentages exceeding 90\%. The transcontinental Eurasian region (comprising Russia, Turkey, Azerbaijan, Georgia, and Kazakhstan) is fully covered. The only underrepresented region is Asia, where missing data for China and India lead to a coverage rate of 60\%. The sample accounts for 87.10\% of the total global land area.

Our analysis is conducted at the country level, and therefore the absence of major states such as China and India does not directly affect the estimation. Nevertheless, it is important to consider the behavior of regions adjacent to these two countries, which rank first and third among the largest emitters of greenhouse gases-widely recognized as the primary anthropogenic driver of climate change-accounting for 32.88\% and 6.99\% of global emissions, respectively (the United States ranks second with 12.60\%).\footnote{Emission data are available at {https://www.worldometers.info/co2-emissions/}.}

\section{ Clusters of Countries with Similar Temperature Changes}\label{sec:cluster}
The first step of our analysis involves clustering the 168 countries based on several criteria derived from their temperature dynamics over 122 years. The characteristics of interest are the rate of warming (hereafter, Clustering A), the temperature variations (Clustering B), and the persistence of the sign of these variations (Clustering C). These features can be quantified using three different statistical tools: the slope of an estimated linear trend, the first-difference operator, and the sequences of signs. The three approaches are presented in the following subsections, along with the corresponding clustering results.

\subsection{Clustering A: Similar Annual Warming Rates}\label{sec:slope}
\begin{figure}[t]
	\centering
	\includegraphics[width=14cm, height=11cm]{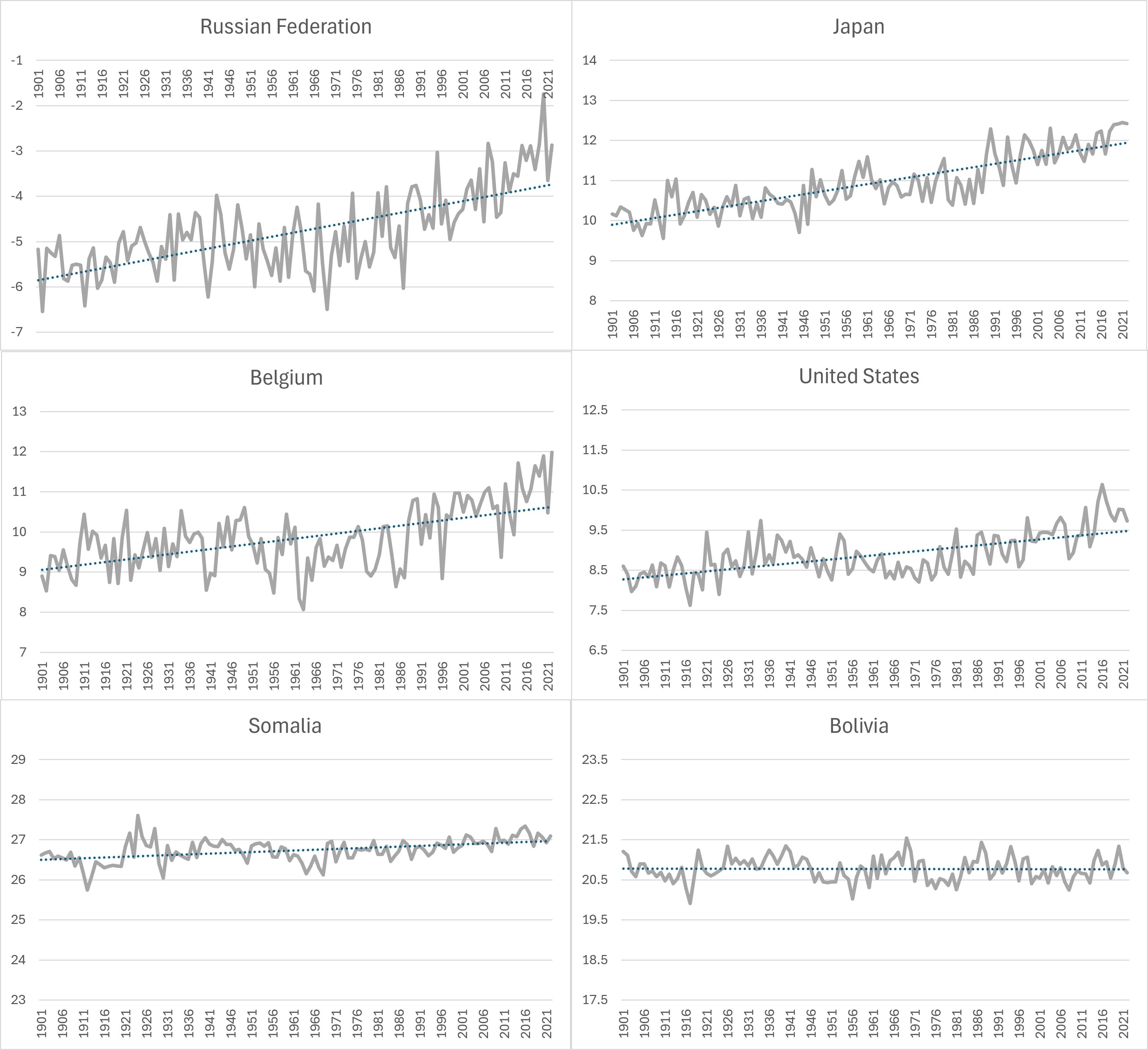}
	\caption{Temperature time series (solid gray lines) of six countries and their corresponding linear trends (dotted black lines).}
	\label{fig:slope}
\end{figure}

The 168 time series display a clear linear trend. Figure~\ref{fig:slope} illustrates the temperature dynamics of six countries, where the dotted line represents the estimated linear trend.\footnote{For comparison, the same range on the $y$-axis (temperature) is set to 6$^\circ$C.} The increase in temperature is approximately linear, but the slopes differ substantially: the Russian Federation and Japan exhibit steep slopes, reflecting a faster rate of warming; Belgium shows a moderate slope; the United States a milder one; while Somalia and, especially, Bolivia display almost flat trends.

The first clustering procedure identifies groups of countries with similar trend slopes. For each country, we estimate a simple linear regression model, where $y_{i,t}$ denotes the temperature of the $i$-th country in year $t$ ($i=1,\dots,168$; $t=1,\dots,122$):
\begin{equation}
	y_{i,t}=a_i+b_i t+\varepsilon_{i,t} \label{eq:linreg}
\end{equation}
where $a_i$ is the intercept, $b_i$ is the slope parameter (interpreted as the annual warming rate), and $\varepsilon_{i,t}$ are zero-mean homoskedastic and uncorrelated errors.  
The distance between countries $i$ and $j$ is computed as the classical Euclidean distance between the estimated slope coefficients, which in this case has the closed form:
\begin {equation}
d_{ij}^b=|b_i-b_j|  \label{dist_b}
\end{equation}
An agglomerative clustering algorithm with the average linkage criterion is then applied \citep[see, e.g.,][]{YANG2017}. The number of clusters is determined from the dendrogram, based on the distance at which clusters merge.

In six cases, the slope coefficient $b_i$ is not significantly different from zero at the 5\% level, suggesting no temperature increase during the observed period. These countries are Bolivia (South America; see Figure~\ref{fig:slope}), Timor-Leste (Asia), Madagascar (Africa), and Kiribati, Nauru, and the Solomon Islands (Oceania). They are excluded from the clustering, which identifies four groups.

\begin{figure}[t]
\centering
\includegraphics[width=14cm, height=10cm]{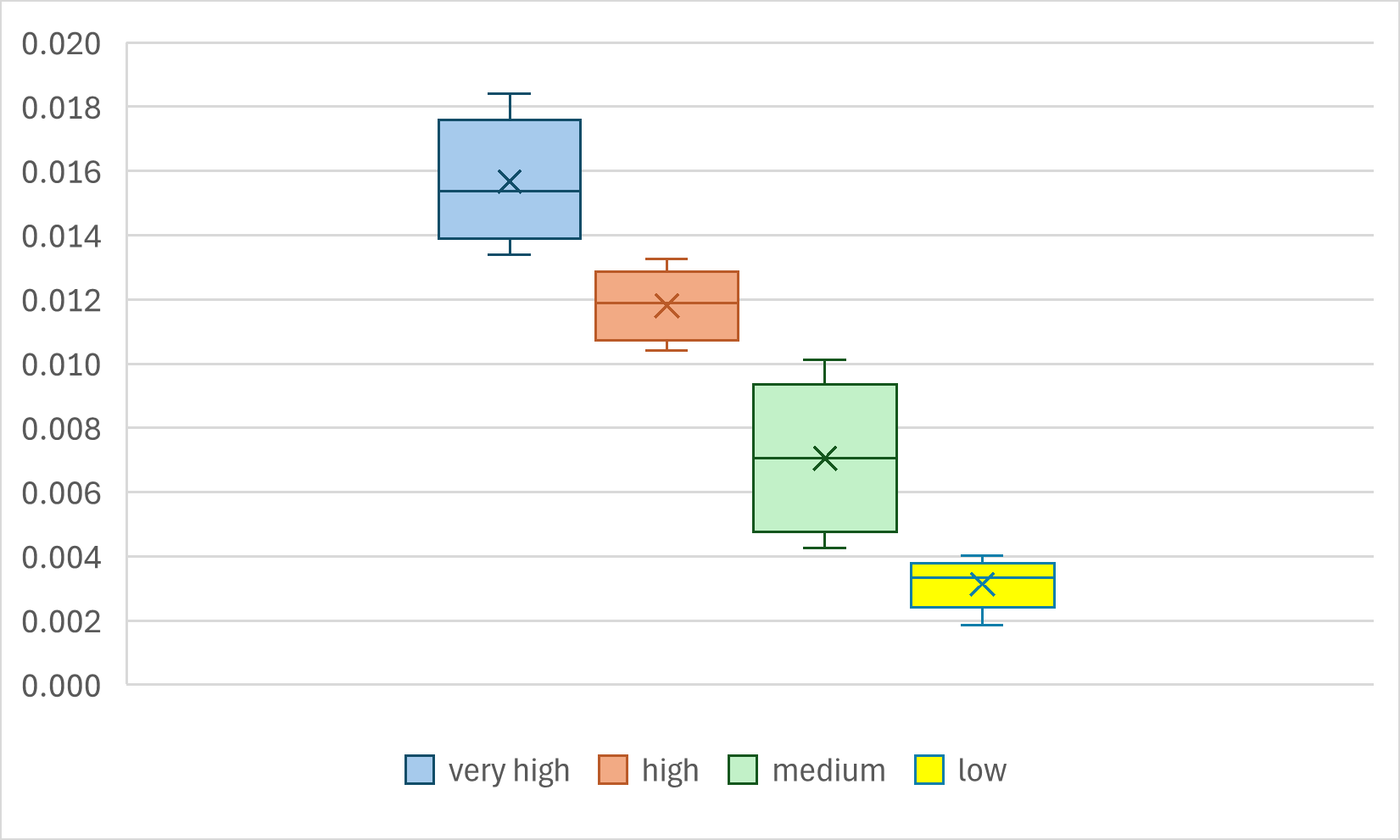}
\caption{Boxplots of the estimated slope coefficients for each cluster. The cross indicates the mean slope within the corresponding cluster.}\label{fig:boxplot}
\end{figure}

Figure~\ref{fig:boxplot} shows that the four clusters are well separated and can be interpreted as very high, high, medium, and low warming rates, respectively.  The very high slope group includes countries with annual warming rates between 0.013 and 0.018 (mean = 0.016); the mean slopes of the remaining clusters are 0.012, 0.007, and 0.003, respectively. The highest rate is observed in Mongolia, likely influenced by its proximity to China.

\begin{figure}[t]
\centering
\includegraphics[width=14cm, height=10cm]{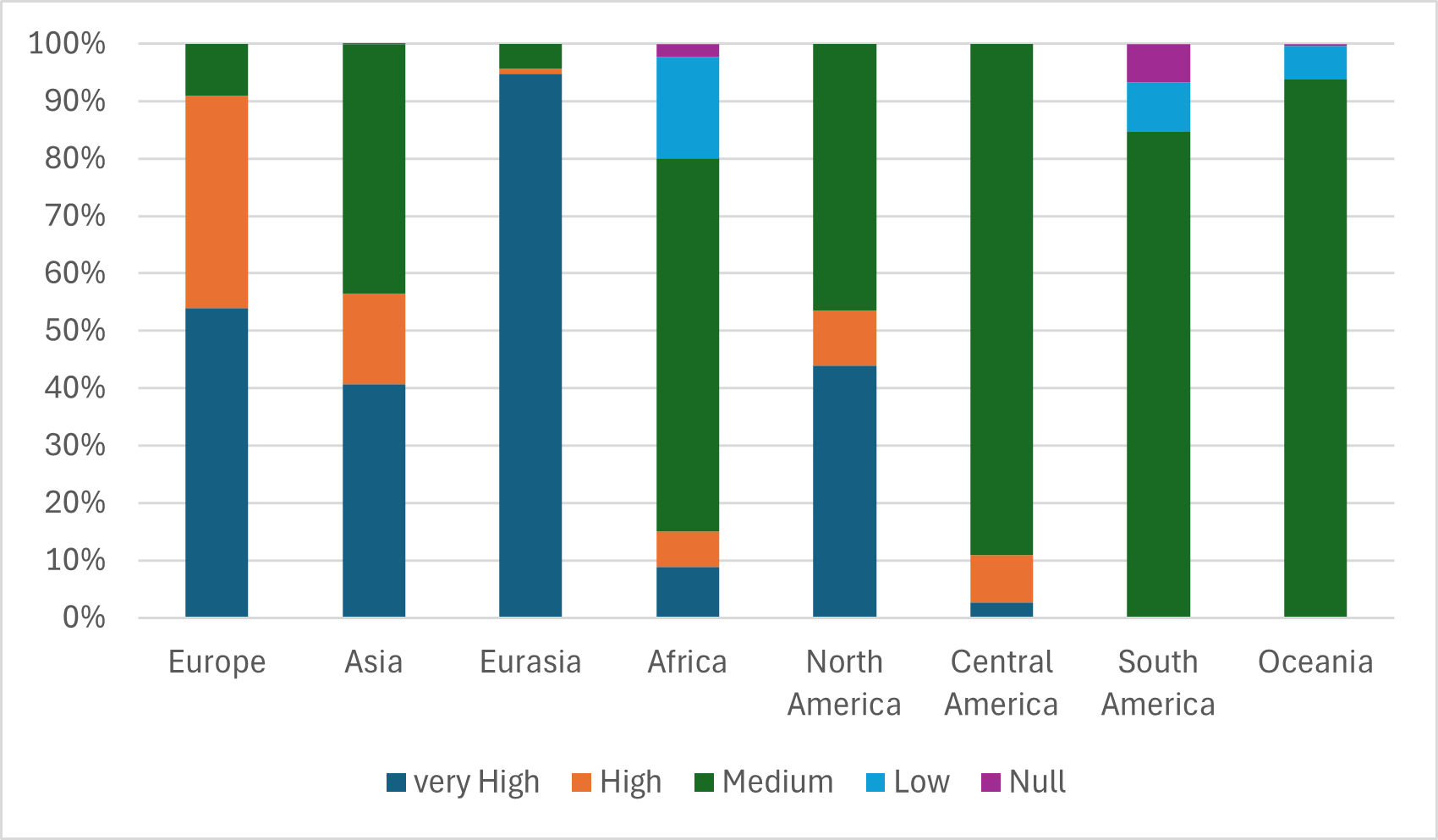}
\caption{Distribution of warming-rate clusters across geographical zones.}\label{fig:geocl1}
\end{figure}

Table~\ref{tab:cl1geo} displays the distribution of countries by geographical region and warming-rate cluster (including the \textit{null} group of six countries with statistically insignificant slopes). European and Asian countries dominate the high and very high clusters, while South America and Oceania exhibit the lowest rates. Weighting countries by land area provides a more accurate picture of territorial influence (Figure~\ref{fig:geocl1}). Eurasia's inclusion of the Russian Federation results in almost complete dominance of the very high cluster in that region. The figure clearly indicates that Europe, Asia, Eurasia, and North America show the strongest warming-consistent with findings on Arctic amplification and greenhouse gas effects-while Africa presents moderate variability and Central/South America and Oceania exhibit lower or even null warming.
A similar observation made for Eurasia holds true for Oceania, where the predominant {\it medium} warming is due to Australia's classification.
\begin{table}[t]
\centering
\begin{tabular}{l r r r r r |r}
	&	very high	&	high	&	medium	&	low	&	null	&		\\
	Europe	&	19	&	14	&	4	&	0	&	0	&	37	\\
	Asia	&	14	&	6	&	17	&	0	&	1	&	38	\\
	Eurasia	&	2	&	2	&	1	&	0	&	0	&	5	\\
	Africa	&	6	&	3	&	29	&	7	&	1	&	46	\\
	North America	&	1	&	1	&	1	&	0	&	0	&	3	\\
	Central America	&	2	&	5	&	7	&	0	&	0	&	14	\\
	South America	&	0	&	1	&	8	&	3	&	1	&	13	\\
	Oceania	&	0	&	1	&	4	&	4	&	3	&	12	\\ \hline
	&	44	&	33	&	71	&	14	&	6	&	168	\\
\end{tabular}
\caption{Distribution of 168 countries by geographical zone and warming-rate cluster.}\label{tab:cl1geo}
\end{table}

\subsection{Clustering B: Similar Temperature Variations}\label{sec:diff}
A complementary clustering approach examines the similarity of annual temperature variations to identify countries experiencing similar fluctuations over time. The goal is to determine whether comparable patterns occur across specific geographical regions.  
Unlike the previous classification, this clustering does not imply a natural order (e.g., very high, high, medium), since the magnitude of temperature variation changes over time: two countries may have similar average variability but differ in yearly dynamics.

We compute first differences of the temperature series using the operator $\Delta$:
\begin{equation*}
\Delta y_{i,t}=y_{i,t}-y_{i,t-1}
\end{equation*}
Clustering is again performed using an agglomerative algorithm with average linkage and the Euclidean distance:
\begin{equation}
d_{i,j}^{\Delta}=\sqrt{\sum_{t=2}^{122} (\Delta y_{i,t}-\Delta y_{j,t})^2} \qquad i,j=1,\dots 168 \label{dist_Delta}
\end{equation}
Five main clusters are identified, while three Arctic-proximate countries (Canada, Iceland, and the Russian Federation) exhibit idiosyncratic behavior and are not assigned to any cluster.

The first cluster (16 countries) is centered in the Middle East, including most of the Arabian Peninsula (except Yemen and Oman) and nearby Eurasian/Asian countries, plus Egypt and Cyprus. The mean temperature change is 0.011 with a standard deviation (SD) of 0.637.  
A second, smaller cluster (5 countries in Central Asia) comprises Uzbekistan, Turkmenistan, Tajikistan, Kazakhstan, and Kyrgyzstan, characterized by higher average variations (mean = 0.018, SD = 0.867).  
Two additional clusters correspond to Europe: a southeastern group (13 countries) with the highest mean variation (0.019, SD = 0.731), and a northwestern/Baltic group (12 countries) with lower mean variation (0.017) but the largest variability (SD = 0.975).  
Finally, a large heterogeneous cluster of 119 countries, labeled \textit{Miscellanea}, shows low average variation (0.007) and low variability (SD = 0.348).

\begin{table}[t]
\centering
\begin{tabular}{l r r r r r |r}
	&\multicolumn{5}{c}{{\bf Clustering A}}\\
	{\bf Clustering B} 	&	very high	&	high	&	medium	&	low	&	null&		\\ 
	SE Europe	&	5	&	6	&	2	&	0	&	0	&13	\\
	NW and Baltic Europe	&	5	&	7	&	0	&	0	&	0&	12	\\
	Middle East	&	7	&	7	&	2	&	0	&	0&	16	\\
	Central Asia	&	4	&	1	&	0	&	0	&	0&	5	\\
	Miscellanea	&	21	&	12	&	66	&	14	&	6	&119	\\
	Idiosyncratic	&	2	&	0	&	1	&	0	&	0	&3	\\ \hline
	&	44	&	33	&	71	&	14	&	6	&168	\\
	
\end{tabular}
\caption{Distribution of 168 countries by Clustering~A (warming rate) and Clustering~B (temperature variation).}\label{tab:clAclB} 
\end{table}
Table~\ref{tab:clAclB} shows the cross-distribution of Clustering~A and Clustering~B. The geographically defined groups in Clustering~B correspond mainly to the medium-to-very-high warming-rate clusters in Clustering~A. Similarly, two of the three idiosyncratic countries in Clustering~B (Canada and Russia) also exhibit very high annual warming rates. The \textit{Miscellanea} group spans all categories in Clustering~A.

\subsection{Clustering C: Persistence of the Sign of Variations}\label{sec:sign}
The magnitude of annual temperature variation can be difficult to compare across countries, as patterns are often heterogeneous; hence the large and geographically mixed \textit{Miscellanea} cluster. An alternative way to capture similarity is to analyze the sign of annual differences (increase or decrease) and their persistence over time.

To measure dissimilarity, we adopt the Hamming distance, a metric introduced by \citet{Hamming:1950} in coding theory to detect and correct errors in programming codes. Denoting $s_{i,t}$ as the $t-$th value (1 or 0) of the string $\bf{s}_i$, representing the sign of the temperature change of country $i$ between years $t-1$ and $t$, the Hamming distance between countries $i$ and $j$ is defined as:
\begin{equation}
d_{ij}^H=\sum_{t=2}^{122} |s_{i,t}-s_{j,t}| \label{dist_H}
\end{equation}
This distance equals the number of differing positions between the binary vectors $\mathbf{s}_i$ and $\mathbf{s}_j$. Clustering is then performed using the standard agglomerative algorithm.

This analysis yields 12 clusters and 20 idiosyncratic countries. Most clusters show clear geographic coherence. The largest includes almost all European countries (34 of 37), except for Malta (which forms a Central Mediterranean cluster with Tunisia), Cyprus (which belongs to the Middle East group), and Iceland (idiosyncratic).  
Four clusters correspond to Asia: the Middle East and Caucasus (18 countries), Southeast Asia (13), Central Asia (8), and East Asia (Japan and South Korea).  
In Africa, 42 of 46 countries belong to two clusters-West Africa (16) and Southern/Eastern Africa (26)-while North Africa exhibits a distinct pattern (e.g., Tunisia and Libya are exceptions).  
In the Americas, northern countries behave idiosyncratically, while most Central and South American nations form a broad cluster of 22 countries; Peru and Ecuador (northwestern South America) form a small cluster, as do Chile and Argentina (southern South America). Uruguay stands alone as idiosyncratic.  
Finally, in Oceania, most countries show independent behavior, with only three Polynesian and Melanesian islands forming a small cluster.

Clustering~C can be viewed as a disaggregation of Clustering~B. As shown in Table~\ref{tab:clBclC}, excluding the \textit{Miscellanea} group, each Clustering~B column corresponds roughly to a single Clustering~C row, while the \textit{Miscellanea} cluster is subdivided among the 12 clusters and idiosyncratic cases of Clustering~C.

\begin{landscape}

\begin{table}[t]
	\centering
	\footnotesize{
		\begin{tabular}{l r r r r r r |r}
			&\multicolumn{6}{c}{{\bf Clustering B}}\\
			{\bf Clustering C}	&	SE Europe	&	NW and Baltic Europe	&	Middle East	&	Central Asia	&	Miscellanea	&	idiosyncratic	&		\\
			Europe	&	13	&	12	&	0	&	0	&	9	&	0	&	34	\\
			Central Mediterranean	&	0	&	0	&	0	&	0	&	2	&	0	&	2	\\
			Middle East and Caucasus	&	0	&	0	&	16	&	0	&	2	&	0	&	18	\\
			Central Asia	&	0	&	0	&	0	&	5	&	3	&	0	&	8	\\
			E Asia	&	0	&	0	&	0	&	0	&	2	&	0	&	2	\\
			SE Asia	&	0	&	0	&	0	&	0	&	13	&	0	&	13	\\
			SE Africa	&	0	&	0	&	0	&	0	&	26	&	0	&	26	\\
			W Africa	&	0	&	0	&	0	&	0	&	16	&	0	&	16	\\
			Central and South America	&	0	&	0	&	0	&	0	&	22	&	0	&	22	\\
			NW South America	&	0	&	0	&	0	&	0	&	2	&	0	&	2	\\
			S South America	&	0	&	0	&	0	&	0	&	2	&	0	&	2	\\
			Polynesia and Melanesia	&	0	&	0	&	0	&	0	&	3	&	0	&	3	\\
			idiosyncratic	&	0	&	0	&	0	&	0	&	17	&	3	&	20	\\ \hline
			&	13	&	12	&	16	&	5	&	119	&	3	&	168	\\
	\end{tabular}}
	\caption{Distribution of 168 countries by Clustering~B (temperature variation) and Clustering~C (persistence of the sign of variation).}
	\label{tab:clBclC} 
\end{table}
\end{landscape}

\section{STAR Models with Distance-Based Weights}\label{sec:star}
A STAR model incorporates both temporal and spatial dimensions within its autoregressive terms. It is among the most widely used frameworks for analysing time series associated with spatial units, as their geographical interdependence affects the dynamics of the individual series. 

Let $x$ denote a stationary variable observed over $N$ spatial units and $T$ time periods. In its simplest and most common specification, considering one lag in both dimensions, the model can be written as:
\begin{equation}
	x_{i,t}=c_i+\phi_i x_{i,t-1}+ \psi_i \sum_{j=1}^N w_{ij} x_{j,t-1}+\varepsilon_{i,t} \qquad i=1,\dots,N; t=1,\dots,T \label{star}
\end{equation}
where $c_i$ is a constant, $\phi_i$ the temporal autoregressive coefficient, and $\psi_i$ the spatial autoregressive coefficient; $\varepsilon_{i,t}$ denotes a zero-mean uncorrelated disturbance. The known values $w_{ij}$ are the elements of a spatial weight matrix, with zeros on the diagonal, representing for each row $i$ the degree of influence of the other spatial units on unit $i$. 

As mentioned, the choice of the spatial weight matrix plays a crucial role in determining the performance of the model. Frequently, contiguity-based weights are employed, assigning a weight of $1/m$ to each of the $m$ immediate neighbours and zero otherwise. However, improved results can often be achieved by deriving the weights from a distance measure directly related to the phenomenon under investigation \citep[see, e.g.,][]{Anselin_Rey:2014}, explicitly defining weights inversely proportional to distance. Given the distance $d_{ij} \le N$ between units $i$ and $j$, the corresponding spatial weights can be derived in two steps. First, the unnormalised weights are computed as:

\begin{equation}
w^*_{ij}=\left\{\begin{array}{cl}
	\frac{N-d_{ij}}{N} & \text{if } i \ne j\\
	0& \text{otherwise}
\end{array}
\right.
\label{wei}
\end{equation}
Then the weights are normalised as:
\begin{equation}
w_{ij}=\frac{w^*_{ij}}{\sum_{j=1}^N w^*_{ij}} \label{norm_wei}
\end{equation}

Given the close connection between the three clustering procedures described earlier and the geographical distribution of countries, we propose to replace $d_{ij}$ in (\ref{wei}) with the distances (\ref{dist_b}), (\ref{dist_Delta}), and (\ref{dist_H}) to obtain the set of spatial weights. This yields six distinct weight matrices:
\begin{itemize}
\item Three matrices are obtained by considering only the distances between countries belonging to the same cluster, assigning zero weights otherwise.\footnote{For example, under Clustering C, Brazil receives a non-zero weight with Bolivia because they belong to the same cluster, while it receives a zero weight with Argentina despite their geographical proximity.} The corresponding STAR models with cluster-based weights are denoted STAR$_{cA}$, STAR$_{cB}$, and STAR$_{cC}$, respectively.
\item Three matrices are obtained by considering all distances (\ref{dist_b}), (\ref{dist_Delta}), and (\ref{dist_H}) and applying equations (\ref{wei})-(\ref{norm_wei}). The corresponding STAR models with distance-based weights are denoted STAR$_{dA}$, STAR$_{dB}$, and STAR$_{dC}$, respectively.
\end{itemize}
Each of the six proposed models is estimated equation by equation using OLS, since the number of spatial units exceeds the number of time periods and a joint estimation would be affected by collinearity issues \citep{Durso_etal:2022}.\footnote{In practice, we assume uncorrelated disturbances across spatial units, such that their relationships are fully captured by the spatial autoregressive component of (\ref{star}).} The time series are made stationary by first differencing ($x_{i,t}=\Delta y_{i,t}$ in equation (\ref{star})). We then evaluate the performance of the models both in-sample and out-of-sample, comparing them with a baseline STAR model whose weights are derived from a contiguity matrix \citep[non-zero weights only for {\it nearest neighbours};][]{Anselin:2013}. This baseline model is denoted STAR$_{NN}$.\footnote{Estimation results are available upon request.}

The in-sample evaluation compares fitted and observed temperatures using the Frobenius norm, which is equivalent to the Mean Squared Error in the univariate case. Denoting fitted temperatures by $\hat{y}_{i,t}$, and letting $\mathbf{Y}=\{y_{i,t}\}$ and $\hat{\mathbf{Y}}=\{\hat{y}_{i,t}\}$ denote the $N \times T$ matrices of observed and fitted temperatures, respectively, the Frobenius norm is defined as:
\begin{equation}
FN=trace\left[(\bf{Y}-\hat{\bf{Y}})'(\bf{Y}-\hat{\bf{Y}})\right] \label{fn}
\end{equation}

\begin{table}[t]
\centering
\begin{tabular}{c r r r r r r r}
	\multicolumn{8}{c}{{\bf In-sample}}\\
	&STAR$_{NN}$&STAR$_{cA}$&STAR$_{cB}$&STAR$_{cC}$&STAR$_{dA}$&STAR$_{dB}$&{\bf STAR$_{dC}$}\\
	FN&4491.5&	4490.6&	4486.8&	4489.2&	4485.8&	4483.2&	{ 4480.3}\\
	\multicolumn{8}{c}{{\bf Out-of-sample}}\\
	&STAR$_{NN}$&STAR$_{cB}$&STAR$_{cC}$&STAR$_{cA}$&STAR$_{dB}$&{\bf STAR$_{dC}$}&{\bf STAR$_{dA}$}\\
	FN&658.3&	656.2&	655.9&	654.1&	654.2&	{ 653.8}&	{ 650.6}\\
	pv&0.000&	0.000&	0.001&	0.001&	0.001&	0.012&1.000\\
\end{tabular}
\caption{Frobenius norm for in-sample and out-of-sample evaluation. For the out-of-sample case, models are ordered from left to right according to the sequence of eliminations in the MCS procedure, down to the best-performing model. The last row reports the associated p-values (pv). Best-performing models are in bold.}\label{tab:inout} 

\end{table}

The upper panel of Table \ref{tab:inout} reports the FN values for the in-sample evaluation. All models using distance-based matrices outperform the baseline model. Among the cluster-based models (subscript $c$), clustering B performs best; among those with distance-based weights (subscript $d$), STAR$_{dC}$ achieves the lowest loss, outperforming all other models.

Statistical climate models are most relevant when assessed in terms of their predictive ability. To this end, we designed the following out-of-sample experiment:
\begin{enumerate}
\item consider the 168 time series from 1901 to 2000;
\item estimate the seven models on this reduced dataset;
\item use the estimated models to forecast the next 22 values;
\item compare observed data for 2001-2022 with forecasts using the $FN$ in (\ref{fn}).
\end{enumerate}

The lower panel of Table \ref{tab:inout} reports the FN values for the out-of-sample evaluation. Again, all distance-based STAR models outperform the baseline. In this case, STAR$_{dA}$ yields the best performance, with an FN value 7.7 points lower than STAR$_{NN}$ and 3.2 points lower than the second-best model (STAR$_{dC}$). 

To formally test whether differences in predictive accuracy are statistically significant, we apply the Model Confidence Set (MCS) procedure \citep{hansen2011mcs}. This iterative testing approach identifies, for a given significance level and loss function, the subset of models with equal predictive ability.\footnote{We refer to \cite{hansen2011mcs} for technical details. The test statistic used here is the semi-quadratic statistic, but the alternative range statistic proposed by the same authors yields identical results. Following their recommendation, the variance of the statistic is obtained using 10,000 bootstrap replications.} At the 1\% significance level, Table \ref{tab:inout} shows that STAR$_{dC}$ belongs to the best-performing set together with STAR$_{dA}$.

In summary, the proposed distance-based STAR models outperform the classical contiguity-based STAR specification both in-sample and out-of-sample. Among them, STAR$_{dC}$-based on the Hamming distance between the signs of temperature variations-achieves the lowest FN loss in-sample and ranks among the best models out-of-sample.

\section{Final Remarks}\label{sec:fin}
Statistical models for capturing and forecasting climate change represent a rapidly growing area of research. In this study, we focus on one of the most widely used space-time models, the STAR model, which is both flexible and straightforward to estimate. A key element in this framework is the spatial weight matrix, which encodes the relationships and mutual influence among countries. In the context of temperature variations, a natural choice is to rely on proximity among neighboring countries. 

Our proposal adopts a data-driven approach, selecting the most similar countries through clustering methods that account for three main characteristics: similarity in annual warming rates, similarity in temperature variations, and similarity in the sign of temperature changes. For the first two characteristics, we employ the classical Euclidean distance to perform clustering, while for the latter, we use the Hamming distance, originally developed in computer science for error detection in coding. After verifying that all three clustering procedures yield clusters that correspond to specific geographic areas, we modify the weight matrix of the classical STAR model by considering either only intra-cluster relationships or the full distance matrices. Among these alternatives, the weight matrix based on the Hamming distance provides the best performance in both in-sample and out-of-sample evaluations.

Clustering has previously been applied in STAR models for various purposes. For instance, \citet{Otranto_Mucciardi:2019} use clustering procedures to reduce the number of unknown coefficients in STAR models, thereby achieving flexible yet parsimonious parameterizations. Similarly, \citet{Durso_etal:2022} apply fuzzy clustering to the parameters of each STAR equation to identify groups of countries with similar dynamics in workplace mobility trends-a procedure conceptually inverse to ours, where clustering is used to define the spatial weight matrix itself.

The primary aim of this work was to propose a novel approach to enhance the fitting and forecasting performance of one of the most widely used models in climate analysis. However, this methodology could be extended to alternative space-time models for climate change \citep[see, e.g.,][]{Kyriakidis_Journel:1999, Le_Zidek:2006}. It may be particularly interesting to apply this approach to other climate-related datasets, not limited to temperature, but including variables such as greenhouse gas emissions, industrialization levels, precipitation, or river flows. Moreover, exploring alternative clustering procedures, such as fuzzy clustering as in \citet{Durso_etal:2022}, represents a promising avenue for further research.

\section*{Funding Sources}
The author gratefully acknowledges the \textit{Italian  PRIN 2022} grant (20223725WE) ``Methodological and computational issues in large-scale  series models for economics and finance".

\appendix
\section{Countries and Clusters}
\label{app1}
In this Appendix, we provide an alphabetical list of the countries included in the dataset, along with their corresponding geographical region (see Table \ref{tab:descr}) and the cluster assignments according to Clustering A, B, and C.

\medskip

\textit{\textbf{Legenda of the clusters}}:

\medskip

{\it Clustering A}: \\
1=medium, 2=high, 3=very high, 4=low, 5=null.

\medskip

{\it Clustering B}:\\
1=Middle East,	2=Central Asia,	3=Southeastern (SE) Europe,	4=Northwestern (NW) and Baltic Europe,	5=Miscellanea,	6=idiosyncratic.

\medskip

{\it Clustering C}:\\
1=Central Mediterranean, 2=northwestern part (NW) of South America, 3=Central Asia, 4=Polynesia and Melanesia,	5=East (E) Asia,	6=southern (S) South America, 7=Europe,	8=Central and South America,	9=Southern and Eastern (SE) Africa, 10=West (W) Africa, 11=Southeast (SE) Asia, 12=Middle East and Caucasus, 13=idiosyncratic.

\centering
\includegraphics{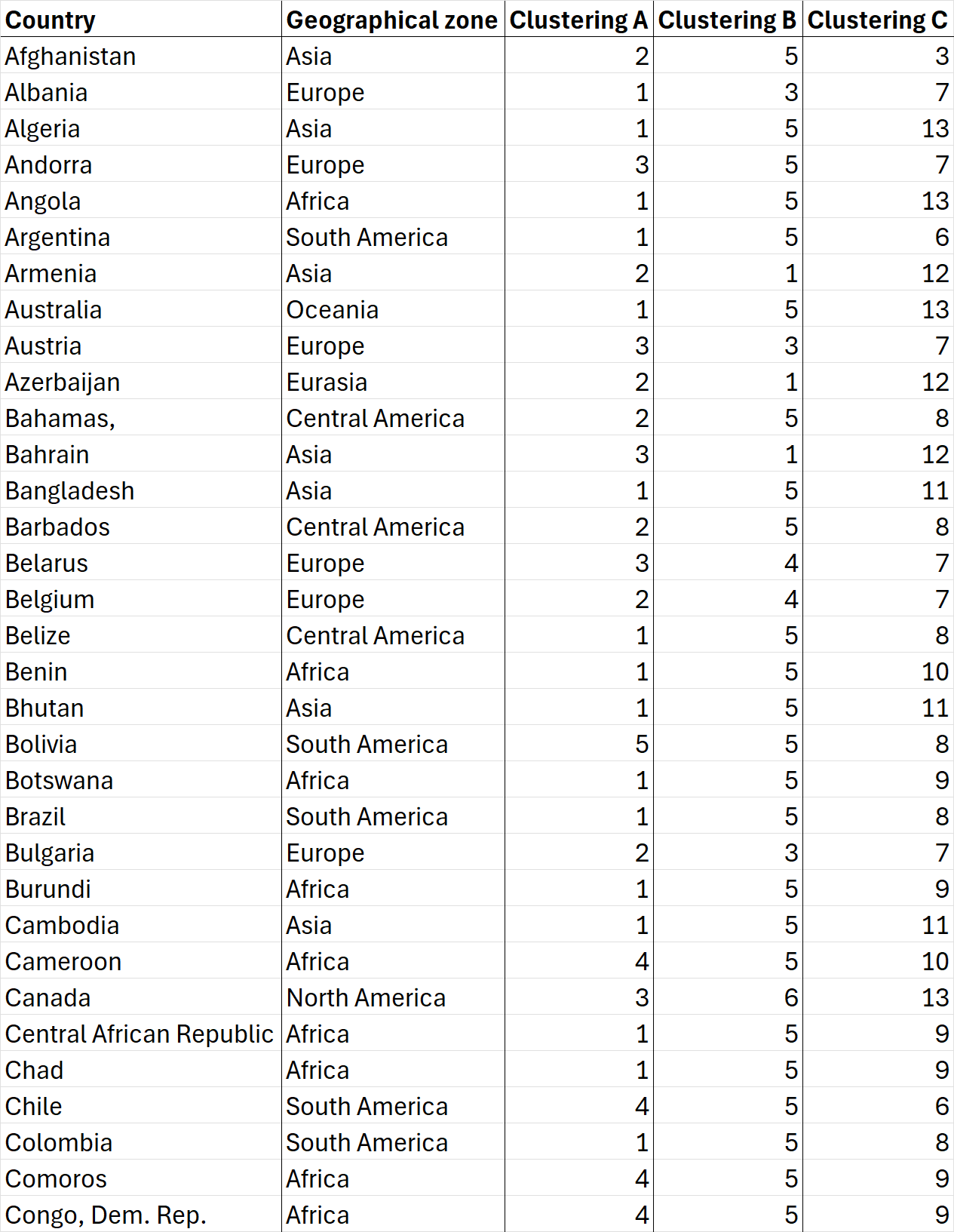}

\newpage

\includegraphics{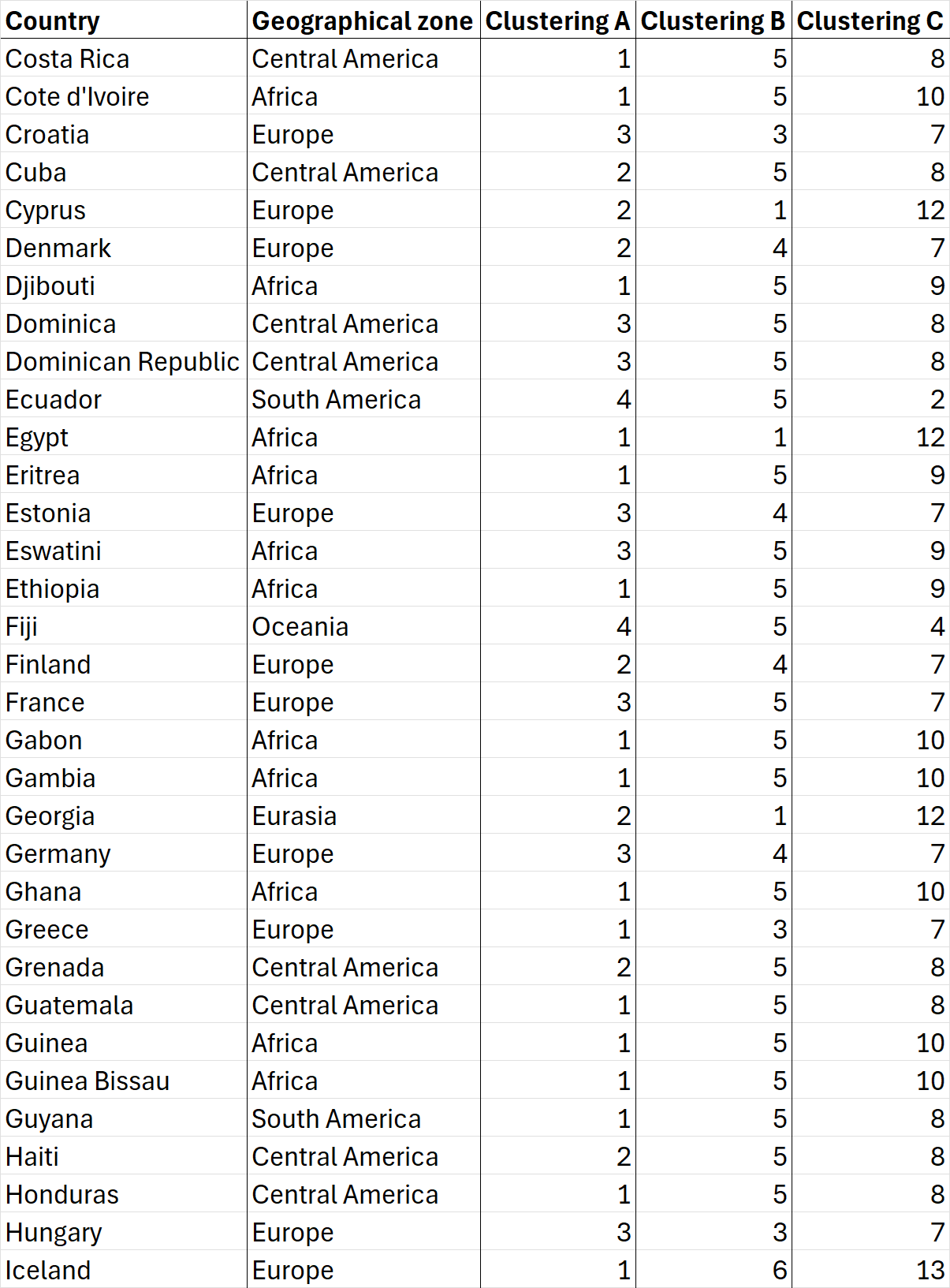}

\newpage

\includegraphics{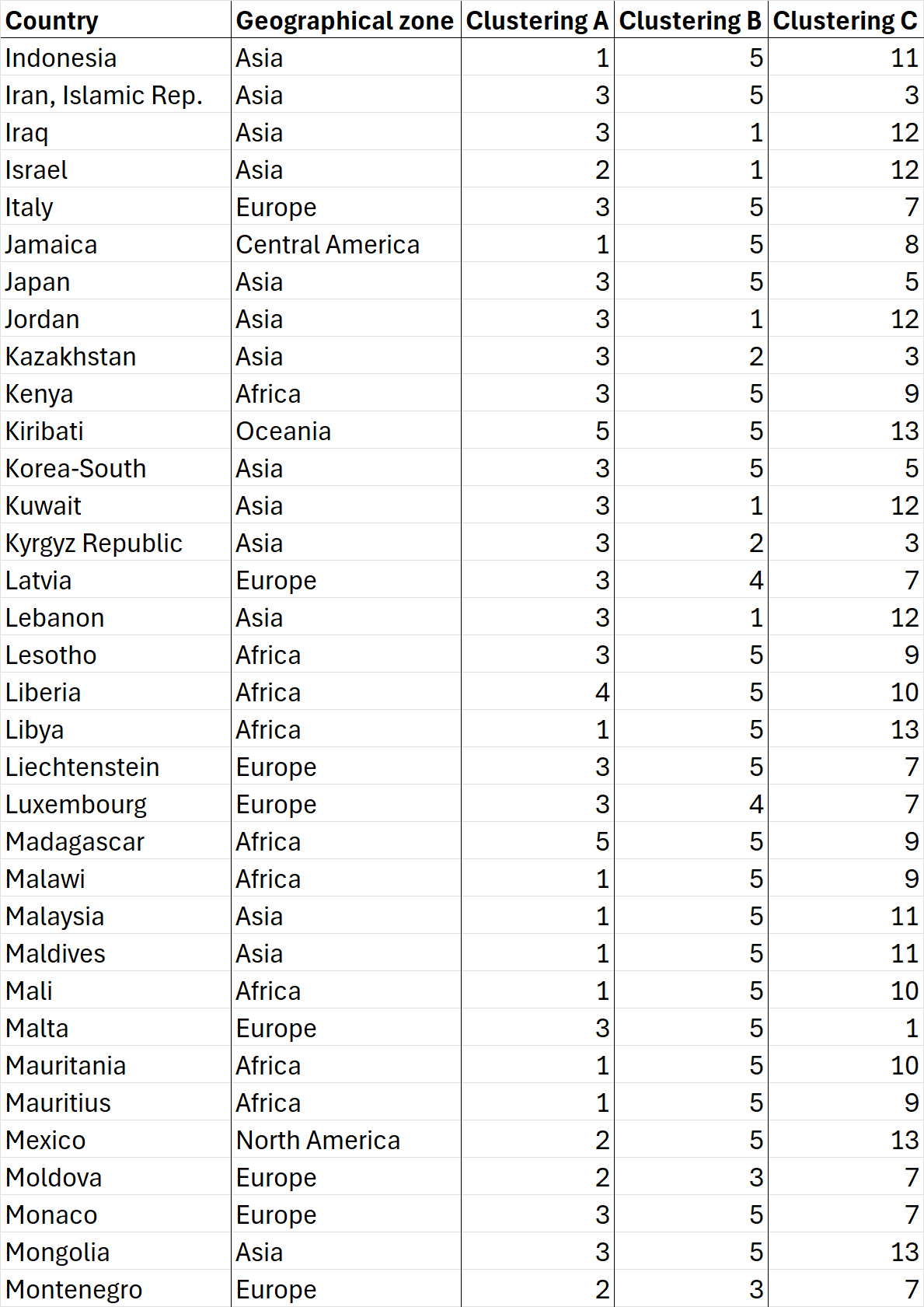}

\newpage

\includegraphics{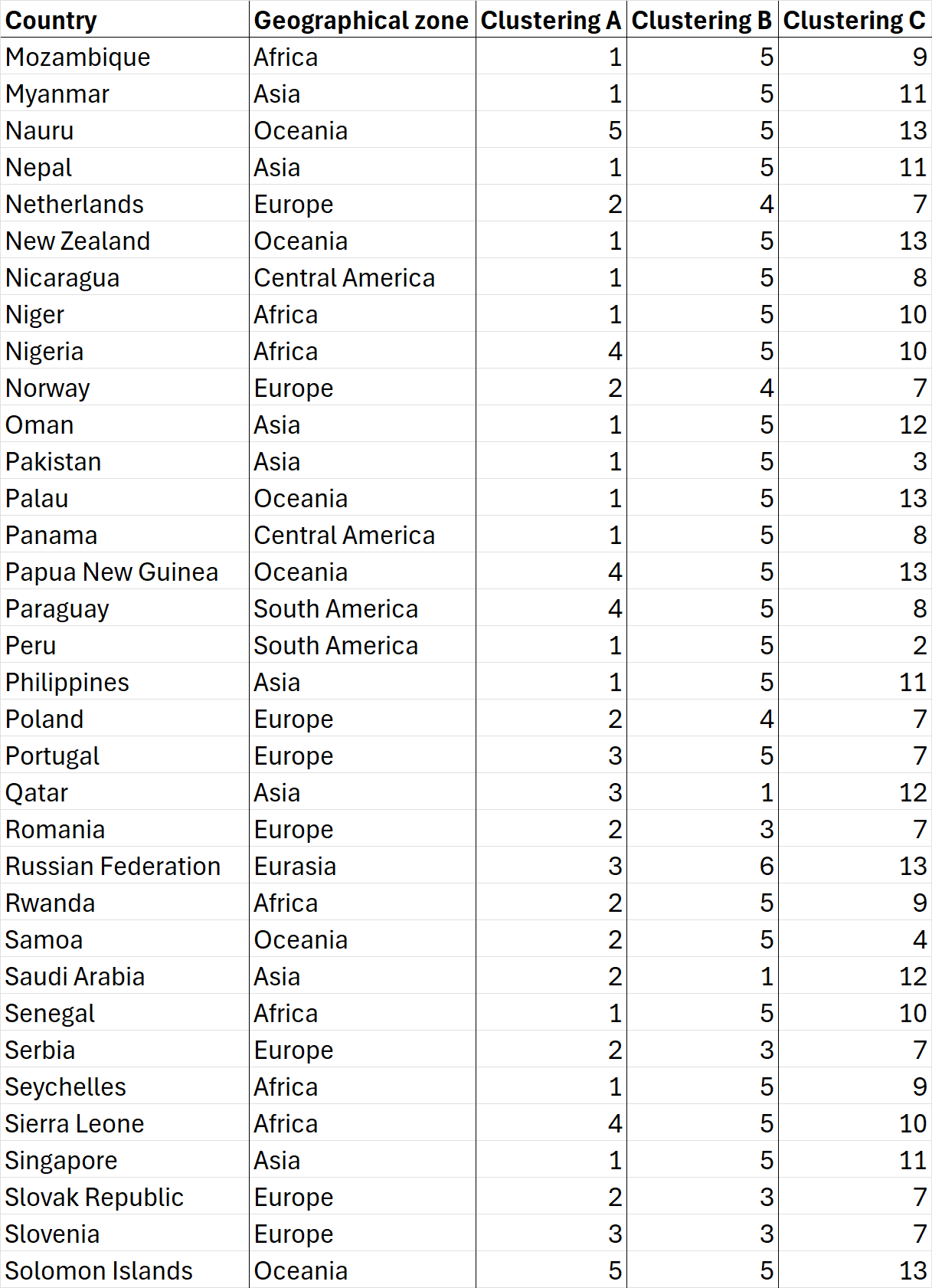}

\newpage

\includegraphics{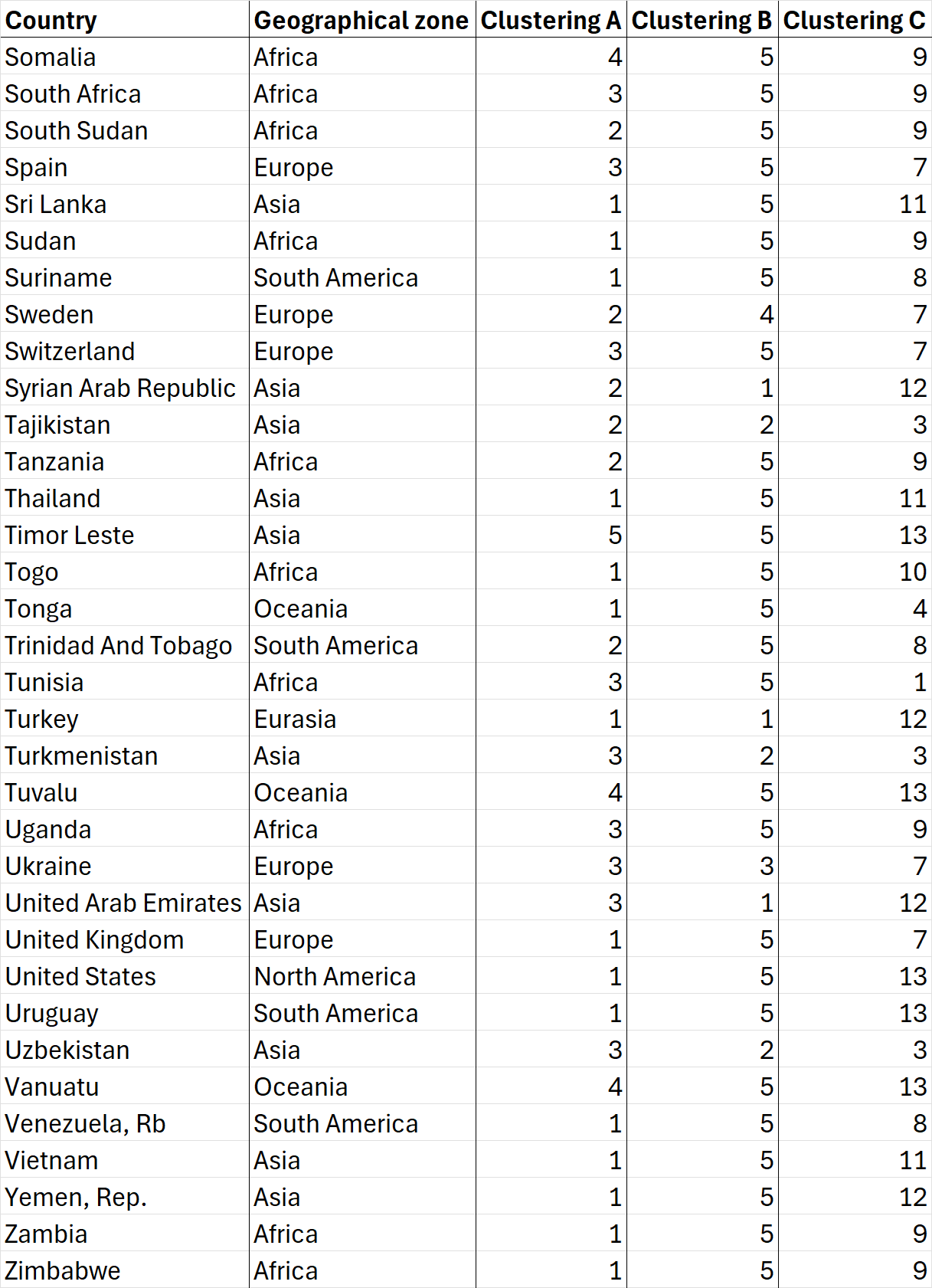}

\end{document}